\documentclass[aps,showpacs]{revtex4}
\usepackage{graphicx}
\newcommand{\beq}{\begin{equation}}
\newcommand{\beqn}{\begin{eqnarray}}
\newcommand{\eeq}{\end{equation}}
\newcommand{\eeqn}{\end{eqnarray}}

\def\H{{u}}
\def\be{\begin{equation}}
\def\ee{\end{equation}}
\def\bea{\begin{eqnarray}}
\def\eea{\end{eqnarray}}

\def\H{{u}}

\begin{document}

\title{Slow-roll corrections to inflaton fluctuations on a brane }

\author{Kazuya Koyama$^1$, Shuntaro Mizuno$^2$ and David Wands$^1$}

\address{
 $^1$Institute of Cosmology \& Gravitation,
University of Portsmouth,
Portsmouth~PO1~2EG, United Kingdom\\
$^2$ Department of Physics, Waseda University, Okubo 3-4-1,
Shinjuku, Tokyo 169-8555, Japan}

\begin{abstract}
Quantum fluctuations of an inflaton field, slow-rolling during
inflation are coupled to metric fluctuations. In conventional four
dimensional cosmology one can calculate the effect of scalar
metric perturbations as slow-roll corrections to the evolution of
a massless free field in de Sitter spacetime. This gives the
well-known first-order corrections to the field perturbations
after horizon-exit. If inflaton fluctuations on a four dimensional
brane embedded in a five dimensional bulk spacetime are studied to
first-order in slow-roll then we recover the usual conserved
curvature perturbation on super-horizon scales. But on small
scales, at high energies, we find that the coupling to the bulk
metric perturbations cannot be neglected, leading to a modified
amplitude of vacuum oscillations on small scales. This is a large
effect which casts doubt on the reliability of the usual
calculation of inflaton fluctuations on the brane neglecting their
gravitational coupling.
\end{abstract}



\maketitle

\section{Introduction}

Inflation is probably the simplest scenario for the origin of
primordial fluctuations in our Universe \cite{inflation}. Small
scale vacuum fluctuations can be stretched to astrophysical scales
by an period of accelerated expansion. Inflation provides a test
of high-energy physics because the perturbations are generated
from very short scales at high energies in the very early
universe. These perturbations carry signatures from high energy
physics, which can be tested by astronomical observations.

The slow-roll approximation \cite{slow-roll} is a useful tool to
study the fluctuations generated during inflation. If we can
neglect the coupling to metric perturbations and the effective
mass of the field then the perturbations are described by the
fluctuations of a free scalar field in de Sitter spacetime. This
gives the familiar result that the power spectrum of scalar field
perturbations at horizon-crossing is given by $(H/2\pi)^2$. One
can then calculate the comoving curvature perturbation which is
conserved on super-horizon scales for adiabatic perturbations.

However inflaton perturbations will be coupled to gravity (metric
perturbations) at first-order in the slow-roll parameters. In
four-dimensional general relativity it is known how to
consistently include linear metric perturbations by working in
terms of the gauge invariant combination of scalar field and
curvature perturbations, the so-called Mukhanov-Sasaki variable,
which obeys a simple wave equation \cite{Mukhanov}. Gravitational
effects are negligible at small scales and high energies, where
perturbations can be normalised to the usual Bunch-Davies vacuum
state. On large scales (super-horizon scales) the comoving
curvature perturbation is conserved allowing one to relate
observations of temperature anisotropies in the cosmic microwave
background to high energy vacuum fluctuations during inflation.
Exact solutions are known for the special case of power-law
inflation in general relativity which generalise the de Sitter
result and have been used to calculate first-order slow-roll
corrections in more general inflation models \cite{S-L}.

In this paper, we develop a new way to derive slow-roll
corrections based on a slow-roll expansion about de Sitter
spacetime. In four-dimensional general relativity we show how to
recover the usual first-order slow-roll corrections. Our method
may be useful when one cannot derive an exact solution and the
background spacetime is given as a perturbation about de Sitter
spacetime.

We then apply our method to inflation in the brane world model.
New ideas in the string theory suggest that our observable
universe is a 4-dimensional hypersurface, or brane, in a higher
dimensional bulk spacetime \cite{Review}. The simplest example of
this model is the Randall-Sundrum model where there is a brane
embedded in a 5-dimensional anti-de Sitter (AdS) spacetime
\cite{RS99}. An AdS spacetime has a characteristic curvature scale
$\mu$ associated with the negative cosmological constant in the
bulk. The spacetime shrinks exponentially away from the brane and
this geometry effectively compactifies the 5-dimensional spacetime
with the effective size $\mu^{-1}$. On large length scales $L >
\mu^{-1}$, 4-dimensional Einstein gravity is recovered, while on
small scales, the gravity becomes 5-dimensional \cite{GT}. In the
early universe when the Hubble horizon is smaller than $\mu^{-1}$,
we expect significant effects from higher dimensional bulk
spacetime. Indeed, the Friedmann equation is modified from the
conventional 4-dimensional theory for $H \mu \gg 1$ \cite{BDEL}.
This modification of Friedmann can provide a novel model for
inflation \cite{MWBH, TMM, binf}.

In Ref.\cite{MWBH}, the amplitude of the curvature perturbation is
calculated by taking into account the modification of the Friedmann
equation. This work has been extended to include higher order
corrections in slow-roll parameters \cite{Liddle} and the formula has
been widely used to confront this model with the observations
\cite{obs}. But to derive these formulae for the spectrum of the
primordial curvature perturbations the effect of coupling to
five-dimensional gravity has been neglected and in particular it is
assumed that the power spectrum of inflaton perturbations at horizon
crossing is given by $(H/2\pi)^2$. This assumption is only valid to
zeroth order in slow-roll parameters. At first order the inflaton
perturbations will be coupled to metric perturbations. In the brane
world, metric perturbations live in the 5-dimensional spacetime, and thus
we must check if 5-dimensional effects change the result of
conventional 4-dimensional theory. Especially, at small scales/high
energies, the 5-dimensional effects could be large.

The first attempt to study the backreaction due to metric
perturbations was made in Ref \cite{KLMW}. There, perturbations
are solved perturbatively in slow-roll parameters. We should
emphasize that this is the only possible way to perform the
calculations analytically. If the background spacetime of the
brane deviates from de Sitter spacetime, we cannot solve the bulk
metric perturbations analytically. In contrast to four-dimensional
general relativity, there are no other exact solutions known for
the perturbation equations. Thus we must develop a new approach to
calculate the effect of slow-roll corrections. In this paper, we
extend earlier studies and investigate the backreaction due to
higher-dimensional perturbations using a slow-roll expansion.

The structure of the rest of the paper is as follows. In section
II, we describe our new approach to derive first order slow-roll
corrections in a conventional 4-dimensional cosmology. In section
III, we review an inflation model in the Randall-Sundrum
brane-world driven by an inflaton field on the brane.  In section
IV, we derive the equations that govern the coupled system of
inflaton fluctuations on the brane and metric perturbations in the
bulk. In section V, the first order corrections to the inflaton
fluctuations on the brane are solved. In section VI, we discuss
the implications of our result for the brane world inflation
model.

\section{Slow-roll expansion of scalar perturbations in 4D cosmology}

\subsection{Background spacetime}

We consider an inflaton $\phi$ whose potential energy density
$V(\phi)$ drives inflation.
In the conventional 4-dimensional general relativity
described by the metric
\begin{equation}
ds^2=-dt^2 + a(t)^2 \delta_{ij} dx^i dx^j,
\end{equation}
the Friedmann equation and the equation of motion for the homogeneous
field, $\phi$, are given by
\begin{eqnarray}
H^2 &=& \frac{\kappa_4^2}{3} \left( \frac{1}{2} \dot{\phi}^2 +V(\phi)
\right), \\
\ddot{\phi} &+& 3 H \dot{\phi} =- \frac{d V}{d \phi},
\end{eqnarray}
where $H=\dot{a}/a$, $\kappa_4=8 \pi G_4$ and $G_4$ is the 4D gravitational
coupling constant. A dot indicates a derivative with respect to
cosmic time, $t$. Slow-roll parameters are defined by
\begin{equation}
\epsilon \equiv - \frac{\dot{H}}{H^2}, \:\:\:\:
\eta \equiv -\frac{\ddot{\phi}}{H \dot{\phi}}.
\label{slowpara}
\end{equation}
Slow-roll inflation is described by small values of
$\epsilon$ and $\eta$.

\subsection{Slow-roll corrections to inflaton fluctuations}

The inhomogeneous inflaton fluctuation, $\delta \phi$, is coupled
to the metric perturbations. In the Longitudinal gauge, the
perturbed metric is written as
\begin{equation}
ds^2=-(1+2 \Psi) dt^2 + a(t)^2 (1+2 \Phi) \delta_{ij} dx^i dx^j.
\end{equation}
The coupled equations for $\delta \phi$, $\Psi$ and $\Phi$
can be simplified by using Mukhanov-Sasaki variable defined by \cite{Mukhanov}
\begin{equation}
u = a \left( \delta \phi - \frac{\dot{\phi}}{H} \Psi  \right).
\end{equation}
Expanding $u$ by Fourier modes, the wave equation
for $u$ is given by
\beqn
\frac{d^2 u_k}{d \tau^2} +
\left( k^2 - \frac{1}{z}\frac{d^2 z}{d \tau^2} \right) u_k =0,
\label{MS}
\eeqn
where $z \equiv (a \dot{\phi})/ H$ and $\tau$ is a conformal time defined as
\begin{equation}
\tau =\int \frac{dt}{a(t)}.
\end{equation}

In the case of the slow-roll inflation, the mass term in
Mukhanov-Sasaki equation (\ref{MS}) can be approximated as
\beqn
\frac{1}{z}\frac{d^2 z}{d \tau^2} = \frac{1}{\tau^2}
\left(2+6\epsilon-3\eta+{\cal{O}}(\eta^2, \epsilon^2)\right),
\label{b_eq_z}
\eeqn
up to first order of the slow-roll parameters. Then
Eq.~(\ref{MS}) can be expressed as
\beqn
\label{sl_Munov_eq}
\frac{d^2 u_k}{d \tau^2} +
\left(k^2 -\frac{1}{\tau^2} (2+6\epsilon-3\eta)\right) u_k =0.
\eeqn
Usually, the appropriately normalized solution with the correct
asymptotic behavior at small scales is obtained by solving
Eq.~(\ref{sl_Munov_eq}) directly as
\beqn
u_k(\tau) = \frac{\sqrt{\pi}}{2} e^{i (\nu+1/2)\pi/2}
(-\tau)^{1/2}H^{(1)}_{\nu}(-k\tau),
\label{sl_Munov_vl}
\eeqn
where $\nu = 3/2 + 2\epsilon -\eta$ and $H^{(1)}_{\nu}$ is
the Hankel function of the first kind of order $\nu$.
Here we assumed the Bunch-Davies vacuum state where perturbations
stay in Minkowski vacuum at small scales.
Equation (\ref{sl_Munov_vl}) is an exact solution of the perturbation
equation (\ref{sl_Munov_eq}) only if the slow-roll parameters
$\epsilon$ and $\eta$ are constant. However their variation in a
Hubble time is second-order and hence of higher-order in the slow-roll
expansion. Thus we can take $\epsilon$ and $\eta$ to be evaluated
around the time of horizon-crossing.

We are interested in the asymptotic form of the solution
well outside the horizon. Taking the limit
$-k \tau \to 0$ yields the asymptotic form of $u_k$;
\beqn
u_k \to e^{i(\nu - 1/2)\pi/2} 2^{\nu-3/2}
\frac{\Gamma (\nu)}{\Gamma (3/2)} \frac{1}{\sqrt{2k}} (-k \tau)^{-\nu+1/2}.
\label{asym_sl_Munov_vl}
\eeqn
Expanding the gamma function in Eq.~(\ref{asym_sl_Munov_vl}), we get
\beqn
u_k \to e^{i (\nu-1/2) \pi/2}
\Big\{1+ (2\epsilon-\eta) (2- \gamma - \ln 2)
\Big\} \frac{1}{\sqrt{2k}}(-k\tau)^{-1-2\epsilon+\eta},
\label{uk_int_of_ah}
\eeqn
where we have used the formula for the poly-Gamma function
\beqn
\psi(3/2) \equiv \frac{\Gamma'(3/2)}{\Gamma (3/2)}
= 2  -\gamma - 2 \ln 2,
\eeqn
where $\gamma$ is an Euler number.

The quantity that is related to observables today is the
the power spectrum of the curvature perturbation given by
\beqn
{\cal{P}}^{1/2}_{\cal{R}} (k) = \sqrt{\frac{k^3}{2 \pi^2}}
\left| \frac{u_k}{z}\right|.
\label{def_curvpert}
\eeqn
From Eqs.~(\ref{b_eq_z}) and (\ref{sl_Munov_eq}), it can be shown that,
at large scale, the time dependences of $u_k$ and $z$ are the same,
that is, ${\cal{P}}^{1/2}_{\cal{R}}$ is constant.
Note that the constancy of ${\cal{R}}$ in the large scale limit does
not depend on the slow-roll approximation, but holds for any adiabatic
perturbation. Thus this comoving curvature perturbation can be related
to the perturbation in the radiation density on large scales long after
inflation has ended.

For the model with a monotonous potential, the following relation holds:
\beqn
|z| = \frac{a |\dot{\phi}|}{H} =
\frac{2}{\kappa_4^2} \frac{a}{H}
\left| \frac{d H}{d \phi}\right|\,,
\label{z_int_of_hphi}
\eeqn
and conformal time can be evaluated up to the first order in slow-roll
parameters as 
\beqn
\tau = -\frac{1}{aH} (1+\epsilon).
\eeqn
Thus the power spectrum of the curvature perturbation is given by
\beqn
{\cal{P}}^{1/2}_{\cal{R}} &=&
[1-(2C + 1) \epsilon + C \eta] \frac{\kappa_4^2}{4 \pi}
 \left\{\frac{H^2}{|dH / d\phi|} \right\}_{k=aH},
\label{s_l_correction}
\eeqn
where $C = -2 + \ln 2 + \gamma \simeq -0.73$.
The terms proportional to the slow-roll parameters are called
Stewart-Lyth correction \cite{S-L}.

\subsection{Perturbing about de Sitter spacetime}

In this subsection, we reproduce the usual slow-roll corrections
in a perturbative approach which does not require any exact solution
of the perturbation equation other than that in a de Sitter spacetime.
This will be more suited to extension to the case of brane-world gravity.

At zeroth order in slow-roll parameters, the spacetime is described by
the de Sitter spacetime. Thus we can expand the spacetime from de
Sitter spacetime. The scale factor is expanded as
\begin{equation}
a(t)=a^{(0)}(t)+a^{(1)}(t) +  {\cal{O}} (\epsilon^2), \quad
a^{(0)}(t)=\exp(Ht).
\end{equation}
Accordingly, the Mukhanov-Sasaki variable is expanded as
\beqn
\label{ex_Munov_vl}
u_k(\tau) = u_k^{(0)}(\tau) + u_k^{(1)} (\tau) + {\cal{O}} (\epsilon^2),
\eeqn
where $u_k^{(0)} \equiv a \delta \phi^{(0)}$ and
$u_k^{(1)} \equiv a (\delta \phi^{(1)} - (\dot{\phi}/H) \Psi)$.
Substituting Eq.~(\ref{ex_Munov_vl}) into Eq.~(\ref{sl_Munov_eq}),
the zeroth order equation is given by
\beqn
\frac{d^2 u_k^{(0)}}{d \tau^2} +
\left(k^2 -\frac{2}{\tau^2}\right) u_k^{(0)} =0.
\label{0th_mukhanov_eq}
\eeqn

Since we expect that the effects of the deviation from de-Sitter
spacetime are insignificant at small scales,
the form of $u_k^{(0)}$ is determined by demanding
a Bunch-Davies vacuum
\beqn
u_k^{(0)}(\tau) = A (-\tau)^{1/2}
H_{3/2}^{(1)}(-k \tau),
\label{0th_Munov_vl}
\eeqn
where $A=(\sqrt{\pi}/2) e^{i \theta}$ and the phase $\theta$
is fixed so that $u_k (\tau) \to (1/\sqrt{2k}) e^{-ik\tau}$.

Next, we must solve $u_k^{(1)}$ sourced by this zeroth order solution
\beqn
\frac{d^2 u_k^{(1)}}{d \tau^2} +
\left(k^2 -\frac{2}{\tau^2}\right) u_k^{(1)}
-\frac{1}{\tau^2}(6\epsilon-3\eta) u_k^{(0)} =0.
\eeqn
If we impose the boundary conditions (i) $u_k^{(1)}(\tau)$
is negligible in the limit $\tau \to - \infty$, and 
(ii) $u_k^{(1)}(\tau)$ does not diverge faster than
$u_k^{(0)}(\tau)$ in the limit $\tau \to 0$,
then we find that the solution is given by,
\beqn
u_k^{(1)} &=&
C_1(-k \tau)^{1/2} J_{3/2} (-k\tau)
+ C_2 (-k \tau)^{1/2} H_{3/2}^{(1)} (-k\tau),\nonumber\\
C_1 &=& \frac{\pi i}{2} (6\epsilon -3\eta) A
\int_{-\infty}^{\tau}d\tau' \frac{1}{\tau'}
\left\{ H_{3/2}^{(1)} (-k\tau')\right\}^2,\nonumber\\
C_2 &=&  \frac{-\pi i}{2} (6\epsilon -3\eta) A
\int_{-\infty}^{\tau}d\tau'\frac{1}{\tau'}
H_{3/2}^{(1)} (-k\tau') J_{3/2}(-k\tau'),
\label{1st_Munov_vl}
\eeqn
where $J_\nu$ is the Bessel function of the order $\nu$.

We take the limit $-k \tau \to 0$ and compare the asymptotic form
with Eq.~(\ref{uk_int_of_ah}). Using the small arguments
limit of the Bessel functions
\beqn
J_{3/2}(x) \sim \left(\frac{x}{2}\right)^{3/2}\frac{1}{\Gamma(5/2)},
\quad
H_{3/2}^{(1)}(x) \sim -i \frac{\Gamma(3/2)}{\pi}
\left(\frac{2}{x}\right)^{3/2},
\label{as_form_bessel}
\eeqn
we can show that the zeroth order Mukhanov-Sasaki variable approaches to
\beqn
u_k^{(0)}(\tau) \to -4i \frac{\Gamma(\frac{3}{2})}{\pi} A
(2k)^{-1/2} (-k \tau)^{-1}.
\label{pert_0th_muk_asa}
\eeqn
Next, we must evaluate the asymptotic form of the first order
Mukhanov-Sasaki variable.
Using Eq.~(\ref{as_form_bessel}), $C_1$ is evaluated as
\beqn
C_1 &\to& 4 i (2\epsilon - \eta) A \frac{\Gamma(3/2)^2}{\pi}
\frac{1}{(-k\tau)^3}.
\label{eq_for_c1}
\eeqn
We should be careful in evaluating the asymptotic behavior of $C_2$
because sub-leading terms are comparable to the contribution from $C_1$.
Using
\beqn
J_{3/2} (x) = \sqrt{\frac{2}{\pi x}}
\left(\frac{\sin x}{x} - \cos x\right), \quad
J_{-3/2} (x) = -\sqrt{\frac{2}{\pi x}}
\left(\sin x + \frac{\cos x}{x}\right),
\eeqn
the integral for $C_2$ in Eq.~(\ref{1st_Munov_vl})
can be evaluated as
\beqn
\int_{- \infty} ^{\tau} d \tau' \frac{1}{\tau'}
H_{3/2} ^{(1)} (-k \tau') J_{3/2} (-k \tau')
\simeq -\frac{2 i}{3 \pi} \mbox{Ci} (-2k\tau) + \frac{14 i}{9 \pi},
\eeqn
where $\mbox{Ci}$ is the integrated cosine function defined as
\beqn
Ci (x) \equiv - \int_x ^{\infty} \frac{\cos t}{t} dt.
\eeqn
For small $-k\tau$, the integrated cosine function can be
expressed as
\beqn
Ci (-2 k \tau) \to \gamma + \ln 2 + \ln (- k \tau).
\label{as_for_int_cos}
\eeqn
Therefore, the asymptotic form of $C_2$ is given by
\beqn
C_2 &\to& (2 \epsilon - \eta) A \left(\gamma + \ln 2 + \ln (- k \tau)
+ \frac{7}{3} \right).
\label{eq_for_c2}
\eeqn
Then we obtain the asymptotic form of $u_k$ for $-k \tau \to 0$
up to the first order in slow-roll parameters
\beqn
u_k (\tau) &\to& (-i) e^{i \theta}
\Big\{ 1+(2 \epsilon - \eta) (2-\gamma-\ln 2) \Big\}
\Big\{ 1-(2 \epsilon - \eta) \ln (- k \tau) \Big\}
\frac{1}{\sqrt{2k}}(-k\tau)^{-1} ,
\label{pert_muk_asb}
\eeqn
where we have used the fact that $A = (\sqrt{\pi}/2) e^{i \theta}$.
This should be compared with Eq.~(\ref{uk_int_of_ah}).
There appears a logarithmic term which diverges for $- k \tau \to 0$.
However, if we can renormalize this divergence by rewriting
the logarithmic term as
\beqn
1-(2 \epsilon - \eta) \ln (- k \tau) \simeq (-k \tau)^{-2\epsilon + \eta}.
\eeqn
we see that Eq.~(\ref{pert_muk_asb}) is consistent with Eq.~(\ref{uk_int_of_ah}).

Indeed, the logarithmic divergence for $-k \tau \to 0$ in $u_k$ does not
show up in the spectrum of the curvature perturbation.
In order to see this, we expand the curvature perturbation as
\beqn
{\cal{P}}_{\cal{R}}^{1/2} =  \{{\cal{P}}_{\cal{R}}^{1/2}\}^{(0)}
+ \{{\cal{P}}_{\cal{R}}^{1/2}\}^{(1)} + {\cal{O}} (\epsilon^2).
\label{expa_curvpert}
\eeqn
On the other hand, by the definition of the curvature perturbation
(\ref{def_curvpert}), we can write the spectrum of
curvature perturbation up to the first order in slow-roll parameters
as
\beqn
{\cal{P}}_{\cal{R}}^{1/2} \simeq
\sqrt{\frac{k^3}{2 \pi}}
\Big| \frac{u_k^{(0)}}{z^{(0)}} + \frac{u_k^{(0)}}{z^{(0)}}
\left(\frac{u_k^{(1)}}{u_k^{(0)}}-\frac{z^{(1)}}{z^{(0)}}\right) \Big|,
\label{expb_curvpert}
\eeqn
where we also expanded $z \equiv (a \dot{\phi})/H$ as
\beqn
z = z^{(0)} + z^{(1)} + {\cal{O}} (\epsilon^2).
\label{exp_z}
\eeqn
Since there is a difficulty to define the curvature perturbation
in de Sitter spacetime,
we concentrate on the ratio between the zeroth order and
the first order of the curvature perturbation.
By comparing Eq.~(\ref{expa_curvpert}) to (\ref{expb_curvpert}),
the ratio is given by
\beqn
\frac{\{{\cal{P}}_{{\cal{R}}}^{1/2}\}^{(1)}}
{\{{\cal{P}}_{{\cal{R}}}^{1/2}\}^{(0)}}
= \frac{u_k^{(1)}}{u_k^{(0)}}-\frac{z^{(1)}}{z^{(0)}}.
\label{ratio_curvpert}
\eeqn
In order to evaluate Eq.~(\ref{ratio_curvpert}), we must obtain
$z^{(0)}$ and $z^{(1)}$, that is, we must solve Eq.~(\ref{b_eq_z})
perturbatively.
Substituting Eq.~(\ref{exp_z}) into Eq.~(\ref{b_eq_z}),
the equation for $z$ at zeroth order is given by
\beqn
\frac{d^2 z^{(0)}}{d \tau^2} = \frac{2}{\tau^2} z^{(0)}.
\eeqn
If we consider only the growing mode,
the zeroth order solution for $z^{(1)}$ can be obtained as
\beqn
z^{(0)} = B \tau^{-1},
\label{0th_zl}
\eeqn
where $B$ is an integration constant.
This zeroth order solution gives a source term in
the equation for $z$ at first order;
\beqn
\frac{d^2 z^{(1)}}{d \tau^2} = \frac{2}{\tau^2} z^{(1)}
+ \frac{(6\epsilon - 3\eta)}{\tau^2} z^{(0)}.
\eeqn
The growing mode solution for the first order $z^{(1)}$
is given by
\beqn
z^{(1)} = -(2\epsilon - \eta) B \tau^{-1} \ln (-k \tau) + BD \tau^{-1},
\eeqn
where $D$ is another integration constant.
Then we get
\beqn
\frac{z^{(1)}}{z^{(0)}} = - (2 \epsilon - \eta) \ln (-k \tau) + D.
\label{ratio_z}
\eeqn
This logarithmic divergence term exactly cancels the logarithmic
divergence term in $u_k$;
\beqn
\frac{u_k^{(1)}}{u_k^{(0)}} =
(2-\gamma-\ln 2 - \ln (-k \tau)) (2\epsilon - \eta).
\label{ratio_mukhanov}
\eeqn
From Eqs.~(\ref{ratio_z}) and (\ref{ratio_mukhanov}) we obtain
\beqn
\frac{\{{\cal{P}}_{{\cal{R}}}^{1/2}\}^{(1)}}
{\{{\cal{P}}_{{\cal{R}}}^{1/2}\}^{(0)}}
= -C(2\epsilon-\eta) -D,
\label{ratio_curvpert2}
\eeqn
where $C = -2 + \ln 2 + \gamma \simeq -0.73$
is again a numerical constant. We cannot determine $D$ in this approach,
which comes from the difficulty to define curvature perturbation
in pure de Sitter spacetime. However, we can still fix $D$ as follows.
Neglecting the logarithmic term, which is canceled by the contribution from $u_k$,
the solution for $z$ is written as
\begin{equation}
z=B (1+D) \tau^{-1}.
\end{equation}
This must be compared with the definition of $z$
\begin{equation}
z=\frac{a \vert \dot{\phi} \vert}{H} \sim -
\frac{\vert \dot{\phi} \vert}{H^2} (1+\epsilon)\tau^{-1},
\end{equation}
where the solution for $a$ up to the first order was used.
Then we can identify $B=-\vert \dot{\phi} \vert/H^2$
and $D=\epsilon$. Then Eq.~(\ref{ratio_curvpert2}) agrees with the
Stewart-Lyth correction given by Eq.~(\ref{s_l_correction}).

\section{Slow-roll inflation in Randall-Sundrum brane world}

In this section, we apply our perturbative approach to the brane-world
model. We consider the simplest version of brane-world inflation model
based on the Randall-Sundrum model. We will consider a single brane
embedded in a 5-dimensional AdS spacetime. We assume that the inflaton
$\phi$ is confined to the brane while gravity can propagate in
the whole 5-dimensional spacetime \cite{MWBH}.

The 5-dimensional metric describing this model is given by \cite{BDEL}
\begin{equation}
ds^2 = dy^2 - N(y,t)^2 dt^2 +A(y,t)^2 \delta_{ij} dx^i dx^j,
\label{BDELg}
\end{equation}
where
\begin{eqnarray}
A(y,t) &=& a(t) \left[ \cosh \mu y -\left(1+ \frac{\kappa^2 \rho}{6 \mu} \right)
\sinh \mu y \right],
\nonumber\\
N(y,t) &=& \cosh \mu y - \left(1- \frac{\kappa^2 \rho}{6 \mu}(2 +3 w) \right)
\sinh \mu y,
\end{eqnarray}
\begin{equation}
\rho=\frac{1}{2} \dot{\phi}^2 +V(\phi), \quad
P= \frac{1}{2} \dot{\phi}^2 - V(\phi),
\end{equation}
and $w=P/\rho$. The brane is located at $y=0$ and the inflaton is
confined to this hypersurface.
On the brane, the Friedmann equation and the equation of motion
for the scalar field are given by
\begin{eqnarray}
H^2 &=& \frac{\kappa_4^2}{3} \rho + \frac{\kappa^4}{36} \rho^2 ,\\
\ddot{\phi} &+& 3 H \dot{\phi} =- \frac{d V}{d \phi},
\end{eqnarray}
where $\kappa_4^2 =\kappa^2 \mu$, $\kappa^2=8 \pi G_5$ and $G_5$ is 5D gravitational 
coupling.
We can define slow-roll parameters in the same way as the
conventional cosmology, 
Eq.~(\ref{slowpara}).

Unfortunately, the background metric (\ref{BDELg}) is not in general a
separable function with respect to $y$ and $t$. Thus we cannot solve
the metric perturbations analytically. In order to solve for the
$y$-dependence of the bulk gravitons and to study the time-dependence
of the perturbations on the brane, we will expand about the special
case of a de Sitter spacetime on the brane. This corresponds to the
background solution to zeroth order in a slow-roll expansion. For a de
Sitter brane, AdS bulk gives a separable form for the bulk
metric~\cite{Kaloper}:
\begin{equation}
 \label{dSbackground}
 ds^2 = dy^2 + N^2(y) \left[ -dt^2 + a^2(t) \delta_{ij} dx^idx^j \right]
  \,,
\end{equation}
where
\begin{eqnarray}
 a(t) &=& e^{Ht} \,,\\
\label{Ay}
 N(y) &=&
 {H\over\mu}\sinh\mu(y_{\rm h}- |y|)\,,
\,.
\end{eqnarray}
and $y=\pm y_{\rm h}$ are Cauchy horizons~\cite{Kaloper}, with
\begin{equation}
y_{\rm h} ={1\over\mu} \coth^{-1}
\left(\sqrt{1+\left(\frac{H}{\mu}\right)^2 }\right)\,.
\end{equation}
It is often useful to work in terms of
the conformal bulk-coordinate $z=\int dy/N(y)$:
\begin{equation}
 \label{defz}
z = {\rm sgn}(y) H_o^{-1}
\ln\left[\coth{\textstyle{1\over2}}\mu(y_{\rm h}-|y|) \right]\,.
\end{equation}
The Cauchy horizon is now at $|z|=\infty$, and the brane is
located at $z=\pm z_{\rm b}$, with
\begin{equation}
z_{\rm b} ={1\over  H}\sinh^{-1}{H\over\mu}\,.
\end{equation}
The line element, Eq.~(\ref{dSbackground}), becomes
\begin{equation}
ds^{2} = N^2(z) \left[ -dt^2 + dz^2 +{\rm e}^{2Ht}d\vec{x}\,^2
\right]\,,
\label{dSmetric}
\end{equation}
where
\begin{equation}
N(z) ={ H\over \mu \sinh(H|z|)}\,.
\label{A}
\end{equation}

\section{Equations for bulk metric perturbations and
inflaton perturbations on the brane}
In this section, we derive the basic equations for the coupled
Mukhanov-Sasaki variable on the brane and bulk metric perturbations
following Ref.\cite{KLMW}.

\subsection{Master variable for perturbations in AdS bulk}
In the background spacetime given by Eq.~(\ref{dSmetric})
bulk metric perturbations can be solved using the master variable \cite{Mukohyama,Kodama}.
The perturbed metric is given by
%
\begin{equation}
ds^2 = N(z)^2 \biggl[ (1+2A_{yy})dz^2
+2 A_y   dt dz
-(1+2 A )dt^2
+ a^2 (1+2{\cal R} )\delta_{ij} dx^i dx^j
\biggr].
\end{equation}
In the special case of a de Sitter brane in the AdS bulk, the
metric variables are written by the master variable $\Omega$ as
%
\begin{eqnarray}
{A} &=& -\frac{a^{-1} N^{-3}}{6} \biggl( 2\Omega''
-3 \frac{N'}{N} \Omega' +{\ddot\Omega} -\mu^2 N^2 \Omega \biggr),
\label{metric_omega_phi}
\\
{A_y} &=& a^{-1} N^{-3} \biggl( {\dot\Omega}' -\frac{N'}{N}{\dot\Omega}
\biggr),
\label{metric_omega_s}
\\
{A_{yy}} &=& \frac{a^{-1} N^{-3}}{6} \biggl( \Omega''
-3 \frac{N'}{N} \Omega' +2{\ddot\Omega} +\mu^2 N^2\Omega \biggr),
\label{metric_omega_n}
\\
{R} &=& \frac{a^{-1} N^{-3}}{6} \biggl( \Omega''
-{\ddot\Omega} -2\mu^2 N^{2} \Omega \biggr).
\label{metric_omega_psi}
\end{eqnarray}
%
From the perturbed 5-dimensional Einstein equation, we can derive
the equation for $\Omega$
%
\begin{eqnarray}
\ddot{\Omega} -3H \dot{\Omega} -\left(\Omega''
-3 \frac{N'}{N} \Omega'\right) + \frac{k^2}{a^2} \Omega
-\mu^2 N^2 \Omega =0.
\label{eq_omega}
\end{eqnarray}
%
Solutions of the master equation can be separated into eigenmodes of
the time-dependent equation on the brane and bulk mode equation:
\[
\Omega(t,y;\vec{x})= \int d^3\vec{k}\, dm\, \alpha_m(t) \H_m(z)
e^{i\vec{k}.\vec{x}} \,,
\]
where
\begin{eqnarray}
\ddot{\alpha}_m -3H\dot{\alpha}_m+\left[ m^2+{k^2\over
a^2}\right] \alpha_m &=&0\,, \label{varphieom}\\
\H_m''-3{N'\over N}\H_m'+ \mu^2 N^2 \H_m + m^2 \H_m &=& 0\,.
\label{bulkmodeeq}
\end{eqnarray}
Note that the Hubble damping term $-3H\dot\alpha_m$ has
the ``wrong sign'', i.e., this is not the standard wave equation
for a scalar field in four-dimensions.

If we write $\alpha_m=a^2\varphi_m$ and
work in terms of the conformal time $\tau=-1/(aH)$, the
time-dependent part of the wave
equation (\ref{varphieom}) can be rewritten as
\[
{d^2 \varphi_m \over d\tau^2}
 + \left[ k^2 - {2-(m^2/H^2) \over \tau^2} \right] \varphi_m = 0 \,.
\]
This is the same form of the time-dependent mode equation commonly
given for a massive scalar field in de Sitter spacetime.
The general solution is given by
\begin{equation}
\label{varphisol}
\varphi_m(\eta;\vec{k}) = \sqrt{-k\tau}\, B_\nu(-k\tau)\,, ~~
\nu^2={9\over4}-{m^2\over H^2}\,,
\end{equation}
where $B_\nu$ is a linear combination of Bessel functions of order
$\nu$. The solutions oscillate at early-times/small-scales for all
$m$, with an approximately constant amplitude while they remain
within the de Sitter event horizon ($k\gg aH$).  `Heavy
modes', with $m>{3\over2}H$, continue to oscillate as they are
stretched to super-horizon scales, but their amplitude rapidly
decays away, $|u_m^2|\propto a^{-3}$. But for `light modes' with
$m<{3\over2}H$, the perturbations become over-damped at
late-times/large-scales ($k\ll aH$), and decay more slowly:
$|u_m^2|\propto a^{2\nu-3}$.

\subsection{Mukhanov-Sasaki equation on the brane}
Now we introduce a scalar field fluctuation on the brane.
We expand the scalar field perturbation in terms of
slow-roll parameters;
%
\begin{equation}
\delta \phi= \delta \phi^{(0)}+\delta \phi^{(1)} +...
\end{equation}
%
The 0-th order of the scalar field fluctuation obeys the following
equation of motion,
%
\begin{equation}
\delta\ddot\phi^{(0)} +3H\delta\dot\phi^{(0)} +\frac{k^2}{a^2}
\delta\phi^{(0)} = 0.
\label{eom_phi0}
\end{equation}
%
The metric perturbations are generated by the 0-th order fluctuation of
the scalar field through the induced Einstein equations on the brane \cite{Deffayet},
%
\begin{eqnarray}
3 H \dot{\Psi}-3 H^2 \Phi + \frac{k^2}{a^2} \Psi
&=& \frac{\kappa_{4,{\rm eff}}^2}{2}
(\dot{\phi} \dot{\delta \phi}_0 +V' \delta \phi^{(0)})
+\frac{\kappa_4^2}{2} \delta\rho_E,
\label{induced_tt}
\\
H\Phi -\dot\Psi &=& \frac{\kappa_{4,{\rm eff}}^2}{2} \dot\phi
 \delta\phi^{(0)} -\frac{\kappa_4^2}{2} \delta q_E,
\label{induced_ti}
\\
-\ddot{\Psi}-3H \dot{\Psi}+H \dot{\Phi}+3 H^2 \Phi
-\frac{1}{3} \frac{k^2}{a^2} (\Psi+\Phi)
&=& \frac{\kappa_{4,{\rm eff}}^2}{2}
(\dot{\phi} \dot{\delta \phi}_0 -V' \delta \phi^{(0)})
+\frac{\kappa_4^2}{6} \delta\rho_E,
\label{induced_ii}
\\
-a^{-2}(\Psi+\Phi) &=& \kappa_4^2 \delta\pi_E,
\label{induced_ij}
\end{eqnarray}
%
where
%
\begin{equation}
A(y=0,t)=\Phi(t),\quad R(y=0,t)=\Psi(t)
\end{equation}
%
%
\begin{eqnarray}
\kappa_4^2 \delta \rho_E &=& \frac{k^4a^{-5}}{3} \Omega \\
 \kappa_4^2 \delta q_E &=& \frac{k^2 a^{-3}}{3}
\left( \dot\Omega -H\Omega \right),
\\
\kappa_4^2 \delta \pi_E &=& \frac{a^{-3}}{2}
\left( \ddot\Omega -H\Omega + \frac{k^2 a^{-2}}{3}\Omega \right),
\end{eqnarray}
%
and
%
\begin{equation}
\kappa_{4,{\rm eff}}= - \kappa_4 \left. \frac{N'}{N} \right
\vert_{y=0}.
\end{equation}
The contributions $\delta \rho_E, \delta q_E$ and $\delta \pi_E$
come from the projected 5D Weyl tensor and these describe the effect
of the bulk gravitational perturbations \cite{SMS}.
The metric fluctuations in turn affect the dynamics of the first order
scalar field perturbation
%
\begin{equation}
\ddot{\delta \phi}^{(1)} + 3H \dot{\delta \phi}^{(1)}
+ \frac{k^2}{a^2} \delta \phi^{(1)}
=-V'' \delta \phi^{(0)} - 3 \dot{\phi}
\dot{\Psi} + \dot{\phi} \dot{\Phi}-2 V' \Phi.
\end{equation}
%

In order to evaluate the effect from metric perturbations,
it is useful to use Mukhanov-Sasaki variable $Q$ as in the conventional
cosmology;
%
\begin{equation}
Q =\delta \phi-\frac{\dot{\phi}}{H} \Psi.
\end{equation}
%
In terms of slow-roll expansion, we have $Q^{(0)} = \delta \phi^{(0)}$ and
$Q^{(1)} = \delta \phi^{(1)} - (\dot\phi/H) \Psi$.
Then using the induced Einstein equations, Eqs.(\ref{induced_tt}),
(\ref{induced_ii}) and (\ref{induced_ij}), we can derive the
equation for $Q^{(1)}$;
%
\begin{equation}
\ddot Q^{(1)} + 3 H \dot Q^{(1)} + \frac{k^2}{a^2} Q^{(1)}
=-V'' Q^{(0)} - 6 \dot{H} Q^{(0)} + J,
\label{Qevolution}
\end{equation}
%
where
%
\begin{eqnarray}
J &=& -\frac{\kappa_4^2 \dot{\phi}}{3 H}
\left(k^2 \delta\pi_E + \delta\rho_E \right) \nonumber\\
&=&-\frac{\dot{\phi}}{H}\frac{k^2 a^{-3}}{6}
\left(\ddot{\Omega}-H \dot{\Omega} + \frac{k^2}{a^2}\Omega \right).
\label{correction_J}
\end{eqnarray}
%
The equation is the same as the standard 4-dimensional cosmology except
for the term $J$, which
describes the corrections from the
5-dimensional bulk perturbations. Because $J$ contains the 5-dimensional
quantity $\Omega$ we must solve the bulk equation for $\Omega$ to
evaluate the effects.

\subsection{Boundary condition for $\Omega$}
In order to solve $\Omega$, we must specify the boundary
condition for $\Omega$.
we rewrite the expressions of $\Phi$ and $\Psi$,
Eq.(\ref{metric_omega_phi}) and (\ref{metric_omega_psi}),
as \cite{HK}
%
\begin{eqnarray}
\Psi &=& \frac{a^{-1}N^{-3}}{6} \biggl[
3 \frac{N'}{N} {\cal F} -3H (\dot\Omega -H \Omega)
-a^{-2} \Delta\Omega \biggr],
\label{psi_fo}
\\
\Phi &=& \frac{a^{-1}N^{-3}}{6} \biggl[
-3 \frac{N'}{N} {\cal F} -3\ddot\Omega +6H \dot\Omega -3H^2 \Omega
+2a^{-2} \Delta\Omega \biggr].
\label{phi_fo}
\end{eqnarray}
%
where
\begin{equation}
{\cal F} = \Omega'- \frac{N'}{N} \Omega.
\end{equation}
Substituting these expressions into the induced Einstein equations
(\ref{induced_tt})-(\ref{induced_ij}),
we obtain the equations written only by $\cal F$ and $\delta \phi^{(0)}$:
%
\begin{eqnarray}
-3 H \dot{\cal F} - k^2 a^{-2} {\cal F}
 &=& \kappa^2 a (\dot{\phi} \dot{\delta \phi^{(0)}} + V'(\phi) \delta
 \phi^{(0)}) \,,
\label{junc_k1b}
\\
\dot{\cal F} &=& \kappa^2 a \dot{\phi} \delta \phi^{(0)} \,,
\label{junc_k2b}
\\
\ddot{\cal F} +2 H \dot{\cal F}
 &=& \kappa^2 a (\dot{\phi} \dot{\delta \phi^{(0)}} -V'(\phi) \delta
 \phi^{(0)})  \,.
\label{junc_k3b}
\end{eqnarray}
%
These equations can be thought as the boundary conditions for
$\Omega$.
Combining the junction conditions, Eqs.(\ref{junc_k1b})-(\ref{junc_k3b}),
we get an evolution equation for ${\cal F}$;
%
\begin{equation}
\ddot{{\cal F}}-  \left(H + 2 \frac{\ddot{\phi}}{\dot{\phi}} \right)
\dot{{\cal F}} + k^2 a^{-2} {\cal F}=0.
\label{eqFb}
\end{equation}
%
This is consistent with the equation for scalar field equation
Eq.(\ref{eom_phi0}).

\section{Perturbative solutions}
We must solve the coupled equations Eqs.~(\ref{eq_omega})
for $\Omega$ and Eq.~(\ref{Qevolution}) for $Q$.
Introducing dimensionless quantities
\begin{equation}
Q(t)=H a(t)^{-1} u(\tau), \quad \Omega(z,t)=\kappa^2 \dot{\phi} H^{-1} \omega(z,\tau),
\end{equation}
the coupled equations are written as
\begin{eqnarray}
k^2 \tau^2 \left( \ddot{\omega} + \frac{4}{\tau} \dot{\omega} + k^2 \omega \right)
&=& \omega'' + 3 \frac{\cosh Hz}{\sinh Hz} \omega' + \frac{1}{\sinh^2 Hz} \omega,\\
\dot{{\cal F}}_{\omega}
&=& a H^2 u, \quad
{\cal F}_{\omega}=\left( \omega' + \frac{\cosh Hz}{\sinh Hz} \omega \right )_{z=z_b},\\
\ddot{u}+ k^2 u - \frac{1}{\tau^2}(2+ 6 \epsilon -3 \eta) u
&=& J_{u}, \quad J_u= -\beta^2 k^2 \tau^2 \left(\ddot{\omega} + \frac{2}{\tau} 
\dot{\omega}
+ k^2 \omega \right),
\end{eqnarray}
where a dot denotes a derivative with respect to $\tau$ and
\begin{equation}
\beta^2 = \frac{\kappa^2 \dot{\phi}^2}{6 H}.
\end{equation}
At the leading order in slow-roll parameters, $\beta^2$ can be
written as
\begin{equation}
\beta^2= \frac{1}{3} \epsilon \frac{H}{\mu}
\left( 1+\left(\frac{H}{\mu} \right)^2 \right)^{-1/2}.
\end{equation}
Thus $\beta^2$ is essentially the slow-rolling parameter and it controls the
strength of coupling between inflaton perturbation and gravitational
perturbations in the bulk. We solve the coupled equations perturbatively
in terms of small $\beta^2$.

\subsection{Zeroth order solutions}
At the zeroth order where $\beta^2=0$, the solution for $u$
is given by
\begin{equation}
u^{(0)}=C_1 (-k \tau)^{1/2} J_{-3/2}(-k \tau)
+C_2 (-k \tau)^{1/2} J_{3/2}(-k \tau).
\end{equation}
Then ${\cal F}_{\omega}$ becomes
%
\begin{eqnarray}
{\cal F(\tau)} = -C_1 H\sqrt{\frac{2}{\pi}} \frac{{\rm cos}(-k\tau)}{-k\tau}
+ C_2 H \sqrt{\frac{2}{\pi}}\frac{{\rm sin}(-k\tau)}{-k\tau}.
\label{F_solution}
\end{eqnarray}
%
This gives the boundary condition for $\omega$.
The solution for $\omega$ in the bulk subject to this condition is obtained as
\cite{KLMW}
%
\begin{eqnarray}
\omega^{(0)}(z,\tau) &=& - 2 C_1
\sum_{\ell=0}^{\infty} (-1)^{\ell}\left(2 \ell+\frac{1}{2} \right)
\frac{(\sinh Hz_b) Q_{2 \ell}(\cosh Hz)}
{\sinh Hz  Q^{1}_{2 \ell}(\cosh Hz_b)}
(-k \tau)^{-3/2} J_{2 \ell+1/2}(-k \tau), \nonumber\\
&& + 2 C_2
\sum_{\ell=0}^{\infty} (-1)^{\ell}\left(2 \ell+\frac{3}{2} \right)
\frac{\sinh Hz_b Q_{2 \ell+1}(\cosh Hz)}
{\sinh Hz Q^{1}_{2 \ell+1}(\cosh Hz_b)}
(-k \tau)^{-3/2} J_{2 \ell+3/2}(-k \tau),
\label{solution_omega}
\end{eqnarray}
%
where the identities
\begin{eqnarray}
\cos(x) &=& \sqrt{2 \pi} \sum^{\infty}_{\ell=0}(-1)^{\ell}
\left(2 \ell + \frac{1}{2} \right) x^{-\frac{1}{2}}
J_{2 \ell+1/2}(x),\label{identity1}\\
\sin(x) &=& \sqrt{2 \pi} \sum^{\infty}_{\ell=0}(-1)^{\ell}
\left(2 \ell + \frac{3}{2} \right) x^{-\frac{1}{2}}
J_{2 \ell+3/2}(x),
\label{identity2}
\end{eqnarray}
were used.

At large scales $-k \tau \to 0$, the dominant contribution
comes from $\ell=0$ mode. On the other hand, on small scales
$-k \tau \to \infty$, all modes becomes comparable and we need to take into
account an infinite ladder of the modes. This means that gravity becomes
5-dimensional at small scales.

\begin{figure}[h]
\centerline{
\includegraphics[width=10cm]{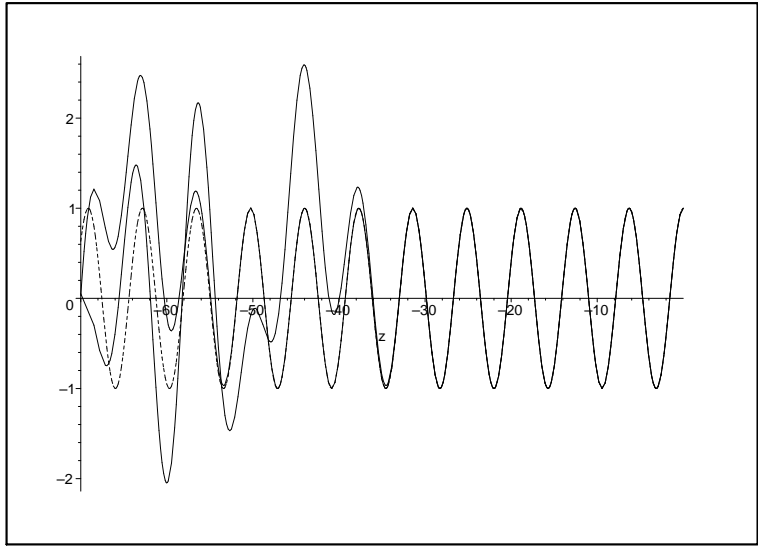}}
\caption{Dotted lines shows $\cos(-z)$ and solid lines show the summation
of Bessel functions with cut-off $\ell_c=20$ and
$\ell_c=30$ respectively. }
\label{fig1}
\end{figure}

In practice, we must approximate the infinite sum to proceed the
calculations. We first check the identity Eqs.~(\ref{identity1}) and
(\ref{identity2}) to see if we can approximate the infinite summation
by introducing a cut-off $\ell_c$ into the summation. {}From
Fig.~\ref{fig1}, we can see that if we increase the cut-off $\ell_c$,
the identity is satisfied for large $-k \eta$, i.e. on small scales.
This implies that as long as we start from a finite time $-k \tau_i$,
we can approximate the infinite ladder of the modes by introducing
sufficiently large $\ell_c$.

\begin{figure}[httb]
\centerline{
\includegraphics[width=13cm]{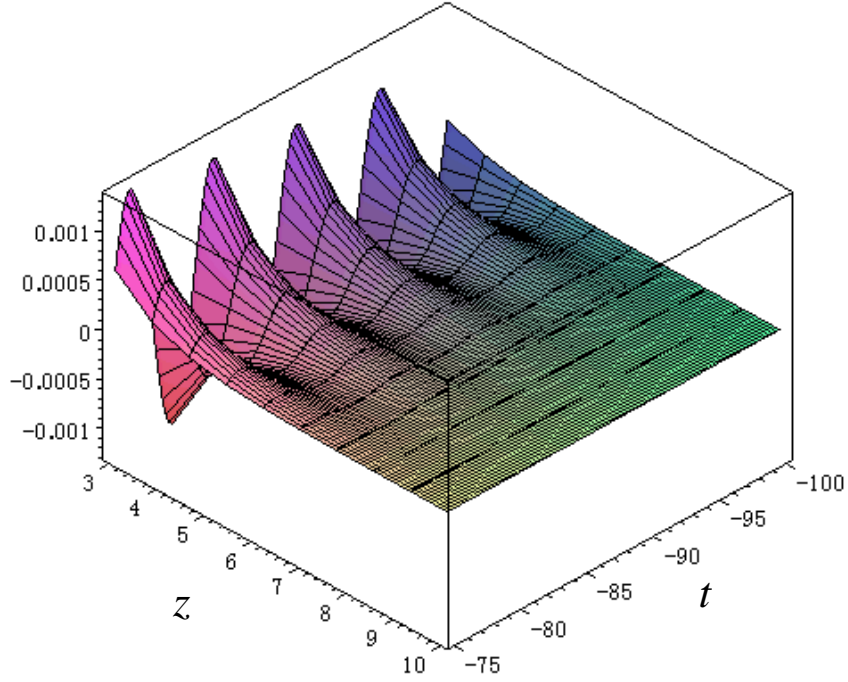}}
\caption{The zeroth order solution for $\omega(z,t)$.
A brane is located at $Hz=3$.}
\label{fig2}
\end{figure}

Fig.~\ref{fig2} shows the bulk solution for $\omega(z,t)$ with
introducing sufficiently large cut-off $\ell_c$. The solution is
localized near the brane and decays towards the horizon $z \to
\infty$. This is a bound state that is supported by an oscillation of
the inflaton fluctuation on the brane. This kind of bound state
generally appears in coupled brane and bulk oscillators \cite{couple}.
A toy example is shown in Appendix. A key point here is that, in this
case, the bound state is a summation of many different eigenstates of
different eigenvalues (Eq.~(\ref{solution_omega})). This fact becomes
crucial in the analysis of the next order solution.

\subsection{First order solutions}

Now it is possible to calculate the next order equation for $u^{(1)}$
\beqn
\label{sl_Munov_eq_rs}
\frac{d^2 u_k}{d \tau^2} +
\left(k^2 -\frac{1}{\tau^2} (2+6\epsilon-3\eta)\right) u_k =J_u,
\eeqn
where $J_u$ describes the effect of the back reaction from the
bulk perturbations.
We can use the 0-th order solution to evaluate
$J_u$ as
\beqn
J_u &=&  \frac{2}{3} \epsilon k^2
C_{1} \sum_{\ell=0}^{\infty} (-1)^{\ell} \left(2\ell+\frac{1}{2} \right)
\bigtriangleup(2\ell;H\mu) \left( 2\ell(2\ell-1) (-k \tau)^{-\frac{3}{2}} 
J_{2\ell+\frac{1}{2}}
+2 (-k \tau)^{-\frac{1}{2}} J_{2\ell+\frac{3}{2}}(-k \tau) \right) \\
&-&  \frac{2}{3} \epsilon k^2
C_{2} \sum_{\ell=0}^{\infty} (-1)^{\ell} \left(2\ell+\frac{3}{2} \right)
\bigtriangleup(2\ell+1;H\mu) \left( 2\ell(2\ell+1) (-k \tau)^{-\frac{3}{2}} 
J_{2\ell+\frac{3}{2}}
+2 (-k \tau)^{-\frac{1}{2}} J_{2\ell+\frac{5}{2}}(-k \tau) \right),
\eeqn
where
\beqn
\bigtriangleup(n;H\mu)= \frac{H}{\mu}
\left( 1+\left(\frac{H}{\mu} \right)^2 \right)^{-1/2}
\frac{Q_{n}(\cosh Hz_b)}
{Q^{1}_{n}(\cosh Hz_b)}.
\eeqn
The quantity $\bigtriangleup(n;H\mu)$ controls the amplitude
of corrections to Mukhanov-Sasaki equations from the bulk
over the change of the energy scales of the inflation.

\begin{figure}[h]
\centerline{
\includegraphics[width=10cm]{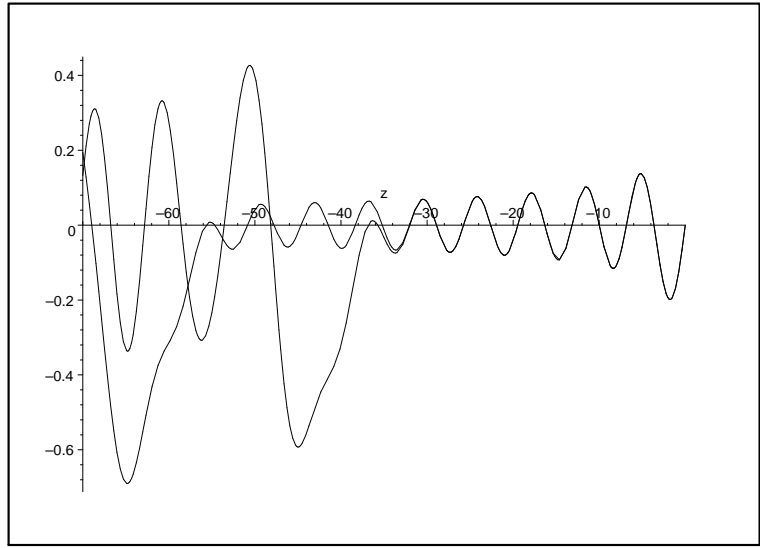}}
\caption{Source term $J_{u}(-z)$ as a function of $z$
with cut-off $\ell_c=20$ and $\ell_c=30$ respectively. Here 
we take $H \mu \gg 1$.}
\label{fig3}
\end{figure}

In order to evaluate $J_u$, we need to introduce a cut-off in the
summation at sufficiently large $\ell$.  Fig.~\ref{fig3} shows the
behaviour of $J_u$ against the change of the cut-off $\ell_c$. A good
feature here is that the behavior of $J_u$ for small $-k \tau$ does
not change even if we increase the cut-off.  Thus we can reproduce the
correct behaviour of $J_u$ by a finite summation of modes as long as
we are considering a finite time interval.

\subsubsection{Large scales}

On large scales $-k \tau \to 0$, $\ell=0$ mode in $C_1$ mode
dominates, which corresponds to a $m^2=2H^2$ mode.  Thus we can
approximate the infinite ladder of the modes by a single mode on super
horizon scales. This indicates that, at large scales, gravity looks
four-dimensional.  Then we can easily show that $J_u$ is suppressed
for $-k \tau \to 0$ and the Mukhanov-Sasaki equation becomes
completely the same as the conventional cosmology. Thus we can show
the conservation of the curvature perturbation ${\cal R}$ on large
scales in the same way as conventional cosmology \cite{LMSW}.

\subsubsection{Small scales}


At low energies $H/\mu \ll 1$, $\bigtriangleup(n;H/\mu)$ can be
approximated as
\begin{equation}
\bigtriangleup(n;H/\mu)=
\left(\frac{H}{\mu}\right)^2 \left(\gamma + \psi(n+1) + \log(H/\mu)- \log 2 \right),
\label{largelow}
\end{equation}
where we assumed $n$ is not large.
Thus, the source terms is well suppressed by the term $\bigtriangleup(n;H/\mu)$
at low energies.
However, at sufficiently small scales, large $\ell$ modes become important
and the approximation (\ref{largelow}) does not hold. Then
we could still get an effect on very sub-horizon scales ($k \gg \mu^{-1} \gg H$).
In this case, we need to introduce a large cut-off in the summation of
$\ell$ and it is technically difficult to perform a calculation.


At high energies, the amplitude of $\bigtriangleup(n;H\mu)$ becomes
large as $H \mu$ becomes large, but, at sufficient high energies
$H\mu \to \infty$, $\bigtriangleup(n;H\mu)$ becomes independent
of $H \mu$ as seen from Fig.4. Indeed, we can obtain the asymptotic
form of $\bigtriangleup(l;H/\mu)$ for $H/\mu \to \infty$ as
\begin{equation}
\bigtriangleup(l;H\mu) \to -\frac{1}{n+1}.
\end{equation}

\begin{figure}[h]
\centerline{
\includegraphics[width=7cm]{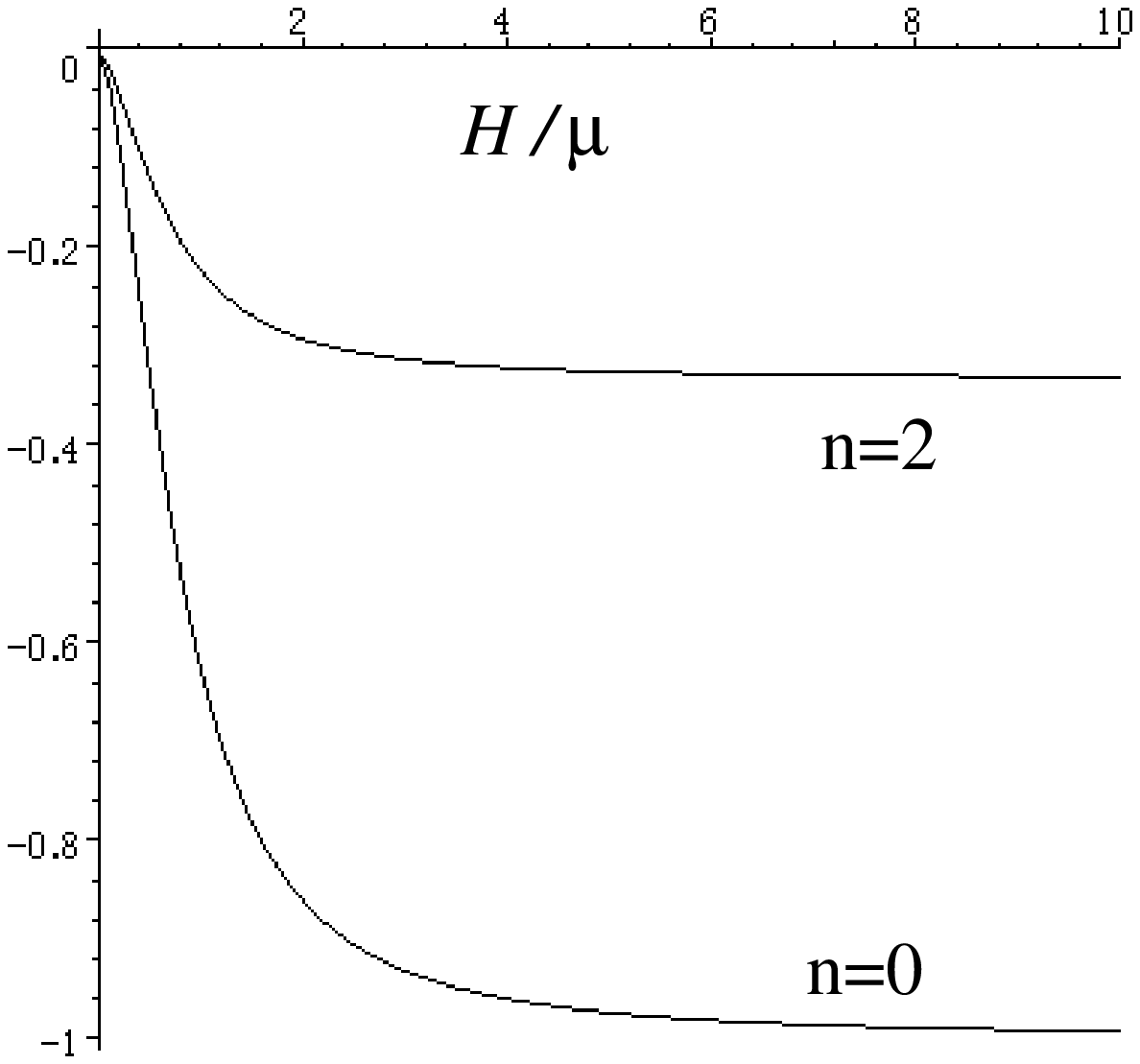}}
\caption{$\bigtriangleup(n;H/\mu)$ as a function of $H/\mu$.}
\end{figure}

In the following, we consider this limit. In this high energy limit,
$J_u$ is well fitted as
\beqn
\label{fitted_ju_rs}
J_u \sim \frac{2 \epsilon}{3} k^2 A \left[   C_1(-k \tau)^{-1/2}
\cos (- k\tau + \varphi) - C_2 (-k \tau)^{-1/2}
\sin (-k \tau + \varphi) \right],
\eeqn
between $-140 < k\tau <-40$ where $A=0.4$ and $\varphi=0.9$.
Then, the equation of motion for the first order Mukhanov variable
is given as
\beqn
\frac{d^2 u_k^{(1)}}{d \tau^2} &+ &
\left(k^2 -\frac{2}{\tau^2}\right) u_k^{(1)}
-\frac{1}{\tau^2}(6\epsilon-3\eta) u_k^{(0)}
- J_u (\tau)=0.
\label{1st_mukhanov_eq_rs}
\eeqn
By using the asymptotic behavior of the Bessel function
at small scale (large $-k \tau$), the third term behaves
like $(- k \tau)^{-2} \sin (- k \tau)$, while the forth term
behaves as $(- k \tau)^{-1/2} \sin (- k \tau)$.
Therefore, at least at small scales, the effect from the bulk
metric perturbations dominates the effect from the standard
corrections to the de Sitter geometry. Thus we will neglect
the third term. The general solutions are given by the linear combination of
$(- k\tau)^{1/2} J_{3/2} (- k\tau)$ and
$(- k\tau)^{1/2} J_{-3/2} (- k\tau)$. By choosing
the initial conditions so that $u_k(\tau_i) = u^{(0)}_k(\tau_i)$ at
$\tau=\tau_i$, we find the following form of the solution,
\beqn
u_k^{(1)} =
D_1(-k\tau)^{1/2} J_{\frac{3}{2}} (-k\tau)
+ D_2 (-k\tau)^{1/2} J_{-\frac{3}{2}} (-k\tau),
\label{1st_Munov_vl_rs}
\eeqn
where $D_1$ and $D_2$ are given by
\beqn
D_1 &=& \frac{\pi}{2} \int_{k \tau_i} ^{k \tau} d (k\tau')
(- k\tau')^{\frac{1}{2}} J_{-\frac{3}{2}} (-k \tau')
J_u (\tau'),\nonumber\\
D_2 &=& -\frac{\pi}{2}
\int_{k \tau_i} ^{k \tau} d (k\tau')
(- k\tau')^{\frac{1}{2}} J_{\frac{3}{2}} (-k \tau')
J_u (\tau').
\label{c1_c2_rs}
\eeqn

For specifying the behavior of the first order Mukhanov variable,
we must evaluate $D_1$ and $D_2$. Using the asymptotic form
for Bessel functions at small scale, $D_1$ and $D_2$ are well approximated as
\beqn
D_1 &\simeq& -\frac{2 \epsilon}{3} A \sqrt{\frac{\pi}{2}}
\int_{k \tau_i} ^{k\tau} d (k\tau')
(- k \tau')^{-\frac{1}{2}} \sin (-k \tau') \left[
C_1 \cos(- k\tau' + \varphi) - C_2 \sin(-k \tau' + \varphi) \right],
\nonumber\\
D_2 &\simeq& \frac{2 \epsilon}{3}  A  \sqrt{\frac{\pi}{2}}
\int_{k \tau_i} ^{k\tau} d (k\tau')
(- k \tau')^{-\frac{1}{2}} \cos (-k \tau')
\left[
C_1 \cos(- k\tau' + \varphi) - C_2 \sin(-k \tau' + \varphi) \right].
\label{c1_c2_rs_sm_as}
\eeqn
Then, on small scales, the first order solution is given by
\begin{eqnarray}
u^{(1)}_k \to
\left((F(\tau)-F(\tau_i) \right) \cos (-k \tau)
+ \left(G(\tau)-G(\tau_i) \right)\sin (-k \tau),
\label{1st_Munov_vl_rs_ap}
\end{eqnarray}
where
\begin{eqnarray}
F(\tau)&=&
\frac{  \epsilon A C_1\sqrt{\pi}}{3}
\left[-S \left(\frac{2 \sqrt{-k\tau}}{\sqrt{\pi}} \right)
\cos \varphi - \left(
C \left(\frac{2 \sqrt{-k\tau}}{\sqrt{\pi}} \right)
-\frac{2}{\sqrt{\pi}} \sqrt{-k \tau} \right) \sin \varphi
\right ] \nonumber\\
&&+
\frac{ \epsilon A C_2\sqrt{\pi}}{3}
\left[S \left(\frac{2 \sqrt{-k\tau}}{\sqrt{\pi}} \right)
\sin \varphi - \left(
C \left(\frac{2 \sqrt{-k\tau}}{\sqrt{\pi}} \right)
-\frac{2}{\sqrt{\pi}} \sqrt{-k \tau} \right) \cos \varphi
\right], \\
G(\tau) &=&
\frac{\epsilon A C_1 \sqrt{\pi}}{3}
\left[ \left( C \left(\frac{2 \sqrt{-k\tau}}{\sqrt{\pi}} \right)
+ \frac{2}{\sqrt{\pi}} \sqrt{-k \tau} \right)
\cos \varphi -
S \left(\frac{2 \sqrt{-k\tau}}{\sqrt{\pi}} \right) \sin \varphi
\right ]\nonumber\\
&&+
\frac{\epsilon A C_2 \sqrt{\pi}}{3}
\left[ -\left( C \left(\frac{2 \sqrt{-k\tau}}{\sqrt{\pi}} \right)
+ \frac{2}{\sqrt{\pi}} \sqrt{-k \tau} \right)
\sin \varphi -
S \left(\frac{2 \sqrt{-k\tau}}{\sqrt{\pi}} \right) \cos \varphi
\right ],
\end{eqnarray}
where $S$ and $C$ are Fresnel functions.

We see that the first order perturbations grows like
$\sqrt{-k \tau}-\sqrt{-k \tau_i}$. Then if we
formally take the limit $-k \tau_i \to \infty$,
the first order corrections diverge. Thus our perturbative
approach breaks down. The amplitude of the zeroth order
oscillation of inflaton fluctuations are significantly affected by
the first order corrections.

We should take care in interpreting this result for the amplitude.
In a toy model of a coupled boundary and bulk oscillators described in
Appendix A, this change of amplitude due to the first order
perturbations is merely caused by the breakdown of the perturbative
expansion. In the toy model, the coupling to the bulk oscillator just
changes the phase of the brane oscillator.  In that case we can
renormalize the first-order perturbation so that the first-order
corrections appear only in the phase of the oscillations and do not
have a large effect on the amplitude.  However, in the case of
inflaton fluctuations, we cannot do this kind of renormalization. This
is due to the phase $\varphi$ in the source term of the first order
equation (see Appendix A).  The phase originates from the fact that
the zeroth order oscillation cannot be matched by a single bulk
eigenmode with the same frequency as the brane oscillator and we need
an infinite ladder of modes. Thus we can say that the effects on the
amplitude from first order corrections are not artificial effects of
our perturbative approach.

In conventional cosmology, the amplitude of inflaton oscillations
$u$ remains constant, so we can impose initial conditions
on any scale far inside the horizon. However, in the brane
world case, the coupling to the bulk metric perturbations
changes the amplitude of the zeroth order inflaton oscillation $u$,
so the effect crucially depends on the initial conditions. In general,
classically, we can also impose arbitrary initial conditions for
$\Omega$. Indeed, it is always possible to add homogeneous solutions which
satisfy the boundary condition given by
\begin{equation}
{\cal F} =0.
\end{equation}
Then we find an infinite tower of massive modes starting from
$m^2 = 9 H^2/4$. Arbitrary initial conditions for $\Omega$
can be satisfied by an appropriate summation of these massive
modes. These massive modes also affect the evolution of
inflaton fluctuations $u$ \cite{HK}.

We have tried to solve the coupled equations for inflaton fluctuations
and master variable directly using a numerical method
\cite{Hiramatsu}.  If we begin with the initial condition for $\omega$
given by Eq.(\ref{solution_omega}), the numerical solution for $u$
well agrees with our perturbative solutions as long as perturbations
remain valid.  We have also tried using different initial conditions for
$\Omega$ and find that the effects on the amplitude of $u$ depend on
the initial conditions for $\omega$ in the bulk.

The initial conditions for $u$ and $\omega$ must be determined by
quantum theory on small scales. Thus we must quantise the coupled
system of the inflaton fluctuations $u$ and the master variable
$\omega$ consistently.  This is in contrast to the conventional
cosmology where we can specify the vacuum for $u$ by neglecting the
gravitational effects far inside the horizon.  This means that the
assumption that the power spectrum of inflaton perturbations at
horizon crossing is given by $(H/2\pi)^2$ could be invalid and we may
have significant effects on the amplitude of perturbations from the
backreaction due to the bulk metric perturbations.

\section{Conclusion}

In this paper we have studied the effect of metric perturbations upon
inflaton fluctuations during inflation, at first-order in slow-roll
parameters $\epsilon$ and $\eta$, which describe the dimensionless
slope and curvature of the potential.
If we neglect the slope and curvature of the inflaton potential then
we obtain the familiar results for free field fluctuations in de
Sitter spacetime, with a scale invariant power spectrum on large
(super-horizon) scales. We take this as our zeroth-order result in a
slow-roll expansion.

In four-dimensional general relativity we were able to calculate
corrections to the field evolution perturbatively to first-order in a
slow-roll expansion, including linear metric perturbations. As far as
we are aware this is the first time the slow-roll corrections have
been calculated in the manner. We reproduce the familiar slow-roll
corrections usually derived from Lyth and Stewart's exact solution to
the linear perturbation equations in power-law inflation.

On a four-dimensional brane-world, embedded in a five-dimensional bulk,
there are no exact solutions for cosmological perturbations (for a vacuum
bulk described by Einstein gravity) except for the case of an exactly de
Sitter brane. Thus the only way to calculate slow-roll corrections is
perturbatively in a slow-roll expansion. We have calculated the leading
order bulk metric perturbations sourced by the zeroth-order inflaton
fluctuations on the brane. We find that inflaton fluctuations support an
infinite tower of discrete bulk perturbations, with negative effective
mass-squared.


Including the effect of the metric perturbations as an inhomogeneous
source term in the wave equation for the first-order inflaton fluctuations
we find that the effect of bulk metric perturbations becomes small on
large scales, and we recover the usual result that the comoving curvature
perturbation becomes constant outside the horizon.

However at small scales (or early times for a given comoving
wavelength) the effect of bulk metric perturbations cannot be
neglected. We are able to give an approximate solution for inflaton
fluctuations at high energies and on sub-horizon scales using a
truncated tower of bulk modes. This shows that the bulk metric
perturbations change the amplitude of inflaton field fluctuations on
the brane. By including a large number of bulk modes we can model this
effect for many oscillations, but ultimately this change of amplitude
becomes a large effect leading to a breakdown of our perturbative
analysis.

It is not surprising in some ways that we see a large effect at small
scales as these are high momentum modes which are expected to be strongly
coupled to the bulk. Nonetheless this invalidates the usual assumption
that gravitational effects are small far inside the cosmological horizon.
It seems necessary to consistently solve for the coupled evolution of
brane and bulk modes. We numerically tried to solve this problem
and verified the validity of our perturbative approach as long as
perturbations remain good. But it was also found that the change of the
amplitude depends on the initial conditions for bulk metric perturbations.
Detailed analysis of numerical solutions go beyond the scope of the present
paper and they will be presented in a separate paper \cite{Hiramatsu}.
In order to give definite predictions for the amplitude of scalar
perturbations in high energy inflation, we must specify the quantum
vacuum state for coupled inflaton fluctuations and metric perturbations
consistently and determine initial conditions. For this purpose,
it would be useful to study the quantum theory of the toy model
for a coupled bulk-brane oscillators in more details where
we can consistently quantise a coupled system \cite{couple}.

Our result implies the possibility that the assumption that the power
spectrum of inflaton perturbations at horizon crossing on a brane is
given by $(H/2\pi)^2$ could be invalid and we may have significant
effects on the amplitude of perturbations from the backreaction due to
the bulk metric perturbations.

\acknowledgements

SM is grateful to the ICG, Portsmouth for their hospitality when this
work was initiated and DW is grateful to the Maeda Lab, Waseda for
their hospitality during its continuation. 
KK is grateful to T. Hiramatsu for the numerical analysis. 
KK is supported by the
Particle Physics and Astronomy Research Council. SM is supported by
the Grant-in-Aid for Scientific Research Fund (Young Scientists (B)
17740154). This work was also supported by PPARC grant PPA/V/S/2001/00544.

\appendix

\section{Toy model for coupled bulk-brane system}

In this appendix, we present a simple toy model for a coupled
brane and bulk oscillators.
Let us consider a toy model for a brane field $q(t)$ and a
bulk field $\phi$ in Minkowski bulk, which satisfy
\begin{eqnarray}
\ddot{q}+\mu^2 q &=& -\beta \phi , \nonumber\\
\ddot{\phi} &=& \phi''-m^2 \phi, \quad \phi'(y=0)= \frac{\beta}{2} q.
\end{eqnarray}
We solve the equations perturbatively in terms of small $\beta$.
Without coupling, the zeroth order solution for $q$ is given by
\begin{equation}
q^{(0)}(t)=C_1 \cos(\mu t) + C_2 \sin(\mu t).
\end{equation}
If we assume $m > \mu$, the 0-th order solution for $\phi$ is obtained as
\begin{equation}
\phi^{(0)}= - \frac{\beta}{2 \sqrt{m^2-\mu^2}} \left( C_1 \cos(\mu t)
+C_2 \sin(\mu t) \right)
e^{-\sqrt{m^2-\mu^2} y}.
\label{solphi}
\end{equation}
Note that the bulk field has a negative effective
mass-squared and decays towards $y \to \infty$.
This is a normalizable bound state supported by an oscillation
of $q(t)$ on the brane.
The equation for the next order $q^{(1)}(t)$ is given by
\begin{equation}
\ddot{q}^{(1)}= - \mu^2 q^{(1)} + \frac{\beta^2}{2 \sqrt{m^2 -\mu^2}}
\left( C_1 \cos \mu t + C_2 \sin \mu t \right).
\label{1stq}
\end{equation}
Including the zeroth order solution, the solution for $q(t)$
is given by
\begin{equation}
q^{(1)}(t)= C_1 \left( \cos \mu t + \frac{\beta^2}{4 \mu \sqrt{m^2 -\mu^2}}
t \sin \mu t\right)
+ C_2 \left( \sin \mu t - \frac{\beta^2}{4 \mu \sqrt{m^2 -\mu^2}}
t \cos \mu t\right).
\label{perturbationq}
\end{equation}
where we impose the initial condition so that $q(0)= q^{(0)}(0)$.

A problem is that the perturbation grows linearly in time.
However, we need to be careful to interpret this
growth of perturbations.
In this toy model, we can easily find an exact solution.
The corresponding exact solution becomes
\begin{eqnarray}
\label{exactq}
q(t) &=& C_1 \cos \left[
\left( \mu -\frac{\beta^2}{4 \mu \sqrt{m^2 -\mu^2}} \right) t
\right]
+ C_2 \sin \left[
\left( \mu -\frac{\beta^2}{4 \mu \sqrt{m^2 -\mu^2}} \right) t
\right]
, \\
\label{exactphi}
\phi(y,t) &=&
-\frac{\beta}{2 \sqrt{m^2 -\mu^2}} q(t) e^{-\sqrt{m^2-\mu^2} y},
\end{eqnarray}
for $\beta \ll 1$. The effect of the coupling merely changes the
frequency of the brane oscillator. The origin of the linear instability
is that the naive expansion in terms of $\beta$ is not
efficient. Indeed, we can use
\begin{eqnarray}
\cos(A+B) &=& \cos A \cos B - \sin A \sin B \sim
\cos A - B \sin A, \nonumber\\
\sin(A+B) &=& \sin A \cos B +\sin B \cos A \sim
\sin A  + B \cos A,
\label{renormalize}
\end{eqnarray}
for $B \ll 1$ and expand the exact solution
into Eq.(\ref{perturbationq}). However, this perturbation breaks down
for large $t$.
A crucial difference of the inflaton fluctuations case
from the toy model is that the source term for the first order equation
for $q$ contains a phase $\varphi$ (compare Eqs.~(\ref{fitted_ju_rs}) and 
(\ref{1st_mukhanov_eq_rs}) to Eq.~(\ref{1stq})).
Then a perturbative solution
cannot be written into the form like Eq.~(\ref{exactq}) using
Eq.~(\ref{renormalize}).
This indicates that there could be a modification of the amplitude
as well as the phase shift. The phase $\varphi$ is originated from the
fact that the brane oscillation cannot be matched by a single
bound state (compare Eq.~(\ref{solution_omega}) and Eq.~(\ref{solphi})).
Thus this is an essential difference between the
toy model and the inflaton fluctuations case.

\end{document}